# GAMMA-RAY BURST ENERGY SPECTRA: THEORETICAL MODELS, OLD AND NEW

MATTHEW G. BARING*

NASA Goddard Space Flight Center, Code 665,
Greenbelt, MD 20771, U.S.A.

**Abstract.** The modelling of gamma-ray burst (GRB) spectra has considerable potential for increasing the understanding of these enigmatic sources. A diversity of ideas and analyses has been generated over the last two decades to explain line features and continuum shapes, encompassing both older galactic neutron star and "new age" cosmological source models. This paper reviews some of the highlights of these studies, discussing the merits and limitations of various ideas, and in particular their compatibility with the observational data. The first focus will be on continuum models for GRBs, which include optically thin synchrotron emission and resonant Compton upscattering near galactic neutron stars, while the synchrotron and non-magnetic inverse Compton scattering mechanisms are prominent in the less well-developed cosmological scenarios. Line formation scenarios will then be discussed, in particular the scattering model for producing cyclotron features, which remains the only viable explanation for the Ginga observations of double lines. Absorption-like line production in cosmological burst models is generally difficult, though interesting notions such as femtolensing interference patterns have been proffered.

## 1. Introduction

Over the years, much effort has been expended in the spectral modelling of gamma-ray burst (GRB) observations. Foremost among these are investigations of line formation: such features often provide powerful discriminating information about the physical conditions and mechanisms operating in emission regions in astronomical sources. Research in this direction was largely focused on describing absorption-like features seen at energies below about 50 keV in some bursts. The observations of GRBs by the KONUS experiment well over a decade ago produced suggestive evidence (Mazets et al. 1981) of single broad absorption features in over 15% of bursts, though the comparatively poor quality of the data recorded at KONUS' low energy range has prompted many astronomers and astrophysicists to be suspicious of the reality of these features. Two to three years prior to the launch of the Compton Gamma-Ray Observatory (CGRO), the Ginga satellite turned up the amazing observation of double absorption-like features in three bursts (Murakami *et al.*, 1988; Fenimore *et al.*, 1988; Yoshida *et al.*, 1991), with the second dip being at twice the energy of the first. At the time, these were interpreted as conclusive proof of cyclotron lines and the "smoking gun" for the galactic neutron star hypothesis of GRBs.

* Compton Fellow, USRA; Email: *Baring@lheavx.gsfc.nasa.gov*

Since the launch of CGRO, neither emission nor absorption lines have been observed with any significance by BATSE (Palmer *et al.*, 1994), confusing the picture and in some quarters creating doubts about the reality of the Ginga line observations. The candidate absorption-like feature in data for one burst seen recently by KONUS aboard the Wind spacecraft (Mazets, 1995) has amplified such confusion. The CGRO-era has emphasized the modelling of GRB continuum spectra, which also has significant, though somewhat lower, diagnostic capability. The detection of prolonged high-energy continuum emission (at $> 100$ MeV) in a significant fraction of bright bursts by EGRET (e.g. see Schneid *et al.*, 1992; Sommer *et al.*, 1994; Hurley *et al.*, 1994) has been a major development in the burst saga: its immediate consequence is the necessity of relativistic motion in these sources (see Baring, 1995a, and references therein). Continuum spectra have been an emphasis of currently popular models of cosmological bursts. Such analyses are generally cruder than galactic neutron star continuum studies, largely because of their youth, though interesting ideas abound, for example the synchrotron and inverse Compton emission predicted from the impact of relativistic "fireball" expansions on external interstellar media. This paper briefly reviews some of the more prominent ideas involved in modelling continuum and line formation, addressing the pros and cons and discriminating features of the various spectral studies.

## 2. Models of Continuum Spectra

Spectral turnovers or breaks are seen in most bursts at a range of energies below about 1 MeV (Band *et al.*, 1993), and the spectral indices either side of them provide crucial information for continuum models. The attenuation of the high-energy GRB continuum by $\gamma\gamma \to e^+e^-$ is an obvious candidate for creating "MeV" turnovers. Inverse Compton scattering of softer radiation can then result from the associated build-up of relativistic pairs and a so-called Compton pair cascade ensues. Such cascades yield flat X-ray spectra typical of GRBs only when producing a broad quasi-thermal bump around 1 MeV (Zdziarski, Coppi and Lamb, 1990) due to pair annihilation and Comptonized thermal bremsstrahlung; this conflicts with the observations. Therefore $\gamma\gamma \to e^+e^-$ must not operate below the maximum energy (i.e. 100 MeV to 18 GeV) detected in EGRET bursts, immediately implying relativistic beaming (e.g. Krolik and Pier, 1991; Baring, 1993) in these bursts in order to blueshift pair production turnovers above the observable energy range. This important result follows because bright bursts are *extremely* radiation-dense if outside the galactic disk, and beaming is a natural way to suppress the $\gamma\gamma \to e^+e^-$ rate due to its angle-dependent threshold. Large bulk Lorentz factors $\Gamma$ are deduced for galactic halo ($\Gamma \gtrsim 3 - 10$) and cosmological ($\Gamma \gtrsim 100 - 10^3$) locations (Harding, 1994; Baring, 1995a), and these may exacerbate GRB occurrence rate problems for various models.

### 2.1. Galactic Neutron Stars

For neutron star models of GRBs, optically thin synchrotron radiation and resonant Compton upscattering (CUSP) provided the strongest candidates for continuum generation. Turnovers around 1 MeV can be generated using synchrotron pair cascades, where magnetic pair production $\gamma \to e^+e^-$ acts to attenuate the continuum (Baring, 1989; Preece and Harding, 1989) since in the strong fields of neutron stars this process dominates $\gamma\gamma \to e^+e^-$ as a means of absorbing gamma-rays. In these cascades, ultrarelativistic electrons are injected into the emission region with significant pitch angles $\theta$ so that a synchrotron continuum is radiated, with the photons being beamed close to the directions of the electrons' momenta. The dramatic increase (e.g. Harding, 1991) in the $\gamma \to e^+e^-$ optical depth above the effective threshold of $2m_ec^2/\sin\theta$ efficiently truncates the continuum above this energy (Baring 1989), subsequently generating a pair cascade. Since this truncation energy is a function of pitch angle, spectra integrated over $\theta$ can generate broken power-laws given suitable angular distributions of electrons (Baring, 1990). The MeV spectral breaks in this scenario can be either substantial or non-existent, an attractive versatility, though synchrotron emission cannot model (Baring, 1993) sources like GRB910503 that have flat spectra (Schneid, et al., 1992) immediately below the turnover.

The inverse Compton scattering of X-rays from the surface of a neutron star by ultrarelativistic electrons can also generate GRB continua, as proposed by Mitrofanov (1984). The resonance of the Compton cross-section at the cyclotron energy $E_{\rm res} = (B/B_{\rm cr})m_ec^2$ in the electron rest frame dominates the scattering process, easily generating spectral breaks (Dermer, 1989, 1990; Baring 1995b, who included $e^-$ cooling self-consistently) at MeV energies (typically $E_{\rm res}^2/\varepsilon_{\rm x}$ for X-rays of energy $\varepsilon_{\rm x}$) when $B \sim 0.1 B_{\rm cr}$. Here $B_{\rm cr} = 4.413\times 10^{13}$ Gauss is the quantum critical field, for which the cyclotron energy is comparable to $m_ec^2$. The spectra are strongly dependent on the viewing perspective, with higher energies being beamed closer to the field lines. Multiple scattering of X-rays should occur if bursts occupy a galactic halo (Dermer, 1990; Baring, 1995b), and has been modelled, omitting $e^-$ cooling, by Ho, Epstein and Fenimore (1992) and Brainerd (1992); they find spectra similar to the single-scattering results of Dermer (1989). Resonant upscattering has no problem dealing with those GRB spectra that are flat below 100 keV, though it has difficulty describing sources with small breaks such as GRB930131, and generating the GeV emission seen in GRB940217.

### 2.2. Continua of Cosmological Bursts

Three interesting ideas for GRB spectral production are discussed here, firstly the concept of Compton attenuation, proposed by Brainerd (1994). If an unspecified cosmological GRB emits power-law gamma-rays and is embedded in a large, dense ($M \sim 10^5 M_\odot$, $r \sim 1$ pc; e.g. molecular) gas cloud, it can Compton scatter radiation out of the line of sight; this occurs preferentially below around 1 MeV because of the decline of the Compton cross-section at higher energies. Hence moderate optical depths yield strong absorption of radiation in the soft gamma-ray regime producing flat spectra below a turnover to a steeper tail, where the cloud is transparent. This nicely generates a characteristic break energy around 1 MeV, though it is difficult to create turnovers much below this (i.e. model all sources) even with the inclusion of cosmological redshifts. Another problem for this scenario is that the attenuation of flux by as much as 100 introduces difficulties in satisfying the extreme energetics of cosmological bursts, if the progenitor is stellar. A variant on this concept, due to Liang (1994), is cold photoelectric absorption by heavy elements like Fe; this dominates the Compton opacity at energies below 200 keV and easily produces X-ray paucity in bursts.

Popular among cosmological burst hypotheses is the *fireball* genre of models. Fireballs can be initiated by the gravitational coalescence of neutron star or neutron star/black hole binaries (e.g. Paczyński, 1986; Eichler *et al.*, 1989), or perhaps partial failure and collapse of a supernova onto its compact core (Woosley, 1993). In either case, roughly a solar mass of energy is released in a very small, optically thick volume, thereby rapidly thermalizing to relativistic temperatures (10 MeV or so, Paczyński, 1986). This energy naturally must disperse adiabatically, and the resulting expansion of baryons and pairs generates relativistic bulk motions, i.e. a fireball. Temperatures around $\sim 20$ keV are usually achieved (e.g. Paczyński, 1986; Goodman, 1986) in the adiabatic cooling of fireballs; rudimentary hydrodynamic calculations (Goodman, 1986) of their emergent spectra yield a cool Planck continuum blueshifted to MeV energies and distorted by bulk motion of the plasma. This is unlike the observed non-thermal spectra of bursts, leading immediately to a problem with pure fireball models, though radiative transfer effects (Carrigan and Katz, 1992) may produce steep high energy tails. Note also that Comptonization of thermal photons by Alfvén turbulence in fireball winds can generate GRB-like spectra (Thompson, 1994).

Fireballs with baryonic content are also relatively inefficient at $\gamma$-ray production (Shemi and Piran, 1990); this property motivated the formulation of "blast-wave" models (e.g. Mészáros and Rees, 1993a,b), where the fireball sweeps up material from the surrounding interstellar medium, creating one or several shocks (much like the propagation of supernova ejecta). The kinetic energy of the fireball is then extracted in non-thermal form via the quasi-isotropic populations of particles accelerated at such shocks. These particles efficiently create non-thermal radiation with broken power-law spectra. Mészáros and Rees (1993b) envisage synchrotron radiation generally in the optical to X-ray range, which for negligible $e^-$ cooling yields breaks of 2/3 or greater in the spectral index at the energy corresponding to the minimum Lorentz factor $\gamma_{\min}$ of accelerated electrons. In the X-ray/$\gamma$-ray range, an inverse Compton "image" of the synchrotron continuum emerges, so that turnovers of suitable severity are predicted. This scenario yields sharper breaks for steeper spectra in the EGRET energy range, which may not be appropriate for all their detected bursts. If cooling is important, the spectral index change at the break diminishes to 1/2 (Mészáros et al., 1994). An important spectral deficiency of this model is that it cannot define a characteristic turnover energy that is restricted to the BATSE energy range.

### 3. Line Features

Before focusing on low-energy GRB absorption features, it is appropriate to discuss briefly the broad emission lines reported at around 400 keV in a small minority of the KONUS data (Mazets et al., 1981). These were naturally interpreted (e.g. Liang, 1986) in the early 80s as 511 keV $e^+e^-$ annihilation lines that are redshifted by the gravitational field of a neutron star. Their widths $\Delta E$ imply temperatures of around 50 keV if attributed to Doppler broadening; however if they originate near a neutron star surface, quantum broadening of such lines in strong magnetic fields becomes important (e.g. Daugherty and Bussard, 1980; Kaminker, Pavlov and Mamradze, 1987). The quantum width of these lines is greatest for emission perpendicular to $\mathbf{B}$, being roughly $\sqrt{B/2B_{\rm cr}}\, m_e c^2$ (e.g. Harding, 1991); only $B \lesssim 0.2 B_{\rm cr}$ can be tolerated by the KONUS widths.

#### 3.1. Cyclotron Scattering Features

It is possible to model the low-energy structure in the KONUS data using the cyclotron emission process. Liang, Jernigan and Rodrigues (1983; see also Brainerd and Lamb, 1987, who included electron cooling) fit various spectra using the dip between the first and second harmonics to mimic an absorption feature. This interpretation yielded dips of relative widths $\Delta E/E \sim 1$, which sufficed for the low-resolution KONUS spectra, but were completely incompatible with the Ginga data, whose features had $\Delta E/E \lesssim 0.25$. Hence attention turned in the late 80s to cyclotron scattering.

The harmonic spacing of the Ginga lines (e.g. at 20 and 40 keV for GRB880205) argued strongly in favour of a "cyclotron absorption" interpretation, an idea voiced years earlier by Bussard and Lamb (1982). While theorists promptly identified true cyclotron absorption $\gamma e \to e^*$ (asterisks denote excited Landau states) as a means of generating such lines in GRBs, Bussard and Lamb (1982, and later Harding and Preece, 1989; see also Ventura et al., 1979, for pulsar contexts) noted that because radiative decay rates are much shorter than collisional rates, cyclotron (resonant Compton) scattering dominates absorption at the first harmonic. Hence, photons are scattered out of the line core instead of being destroyed. Harding and Preece (1989, see also Fenimore et al., 1988, Wang et al., 1989) observed that for photons originally at the $n^{\rm th}$ cyclotron harmonic ($E_n/m_e c^2 \sim nB/B_{\rm cr}$, $n = 1, 2 \ldots$), the scattering events predominantly produce electrons in the $(n-1)^{\rm th}$ Landau level, with scattered photons being at the fundamental (i.e. $n = 1$). The subsequent sequential cyclotronic de-excitations of the electron to the ground Landau state spawns additional cyclotron photons at the fundamental. See Harding (1991) for a review of these processes.

Wang *et al.*, (1989) performed detailed modelling of the Ginga line features in GRB880205 using a Monte Carlo radiative transfer technique to describe the interaction of a power-law continuum spectrum with a strongly-magnetized slab. The remarkable success of their fit to the data for a field of $B = 1.7 \times 10^{12}$ Gauss was due both to the spawning process and also to scattering in the line core. The Compton cross-section at the $n^{\rm th}$ harmonic declines rapidly with increasing $n$ (e.g. Harding and Preece, 1989), so that in a pure absorption scenario, the dips would correspondingly decrease in depth with increasing $n$, probably disappearing (into the noise of the data) beyond the third harmonic. However, since the scattering process spawns many cyclotron photons, the first harmonic is partially filled and in the computations of Wang *et al.*, (1989, see also Alexander and Mészáros, 1989, who included polarization dependence) becomes comparable in strength to the second dip, as predicted by Bussard and Lamb (1982).

The narrow widths of the absorption dips in the Ginga detections implied a cold scattering medium. Such low temperatures were explained by Lamb, Wang and Wasserman (1990) as being due to the equilibrium between heating and cooling of electrons in resonant Compton scattering by a power-law photon continuum; they obtained a so-called "resonant Compton temperature" of $kT \sim 0.25(B/B_{\rm cr})m_e c^2$ ($\sim 5$ keV for $B = 1.7 \times 10^{12}$ Gauss), yielding remarkable consistency between the observed line widths and their energies. The success of these analyses hinged on column densities being around $10^{21}$ cm$^{-2}$ and viewing perspectives being almost perpendicular to the field. The line strength and width is strongly dependent on such a perspective (Graziani *et al.*, 1992), with the fundamental broadening and higher harmonics disappearing when viewed along the field. The temporal variability and disappearance of the dips in the Ginga observations were attributed by Graziani *et al.*, (1992) to neutron star rotation, with Ginga sampling different angles to the field at different times; they inferred a rotation period of between $45s$ and $180s$ from the line variability of GRB870303.

While the cyclotron absorption model is extremely successful, it is difficult to picture how a cold scattering layer can be maintained above a hotter region that produces the continuum. Furthermore, in this scenario bursts must be closer than about 500 parsecs to be sub-Eddington, so that static neutron star environments are totally untenable if bursts originate in the halo of our galaxy, a currently popular hypothesis. Wind scenarios therefore must be envisaged, and Miller *et al.*, (1991) performed limited calculations of higher harmonic cyclotron (i.e. absorption) line formation in relativistic winds emanating from neutron stars, suggesting that line energies and widths similar to the static case could be produced because the absorption occurs in cold regions of lower $B$, but is Doppler-boosted to observed energies. Full radiative transfer and hydrodynamic calculations in such winds including scattering at the fundamental are yet to be performed.

### 3.2. Lines in Cosmological Scenarios

It is difficult to envisage absorption line production in bursts of extragalactic origin, for example relativistically-blueshifted atomic lines, largely because of high ( = ionizing) temperatures expected in source models. A novel notion for line production in cosmological bursts is *femtolensing* of the source by intervening "massive" extended or compact objects. In the so-called geometrical optics limit, appropriate for lenses of galactic scales, lensing is achromatic and images of a source have identical spectra; no evidence for such "macrolensing" of bursts exists (Nemiroff, 1993). However, if the wavelength of a photon becomes comparable to the Schwarzschild radius $R_s = GM/c^2$ of the lens, geometrical optics fails and the wave nature of light must be considered (Deguchi and Watson 1986). Observable effects then include spectral "interference patterns" or distortions, formed by constructive and destructive interference between light waves, and arise if the source is comparable in size to the Einstein radius of the lens. Gould (1992) postulated that cosmological GRBs could satisfy this criterion, and coined the term femtolensing for this phenomenon. Normally such interference patterns are pictured as variations in space, however in this context, since the line of sight is fixed, the light signal is amplified at certain energies and diminished at others. The resulting "fringe pattern" is pronounced when the source is well within the Einstein ring of the lens and has periodic dips at harmonics of the fundamental energy $E = \hbar c / R_s$ (Stanek *et al.*, 1993). This attractive concept fails to explain the Ginga data because it predicts both broad widths ($\Delta E \sim E$) and a second harmonic weaker than the first. It also suffers from there being no particular reason why Nature should select lensing masses of $10^{19}$ g, which are required to match the observed line energies ($E \sim 20$ keV).

This brief, compact summary of spectral studies of bursts gives a flavour of the broad range of interesting ideas from both galactic and cosmological scenarios, and their deficiencies and strengths. Theoretical work on spectra continues to influence experimental agendas, and clearly viable models of absorption-like line mechanisms in cosmological bursts, winds emanating from galactic neutron stars, and ideas for producing prolonged high energy emission are important priorities for future studies. In the meantime, gamma-ray bursts remain as enigmatic as ever.

I thank Peter Mészáros, John Wang, Alice Harding and Martin Rees for providing suggestions aiding the refinement of the manuscript.

### References

Alexander, S. G. and Mészáros, P.: 1989, *Astrophys. J. (Lett.)* **344**, L1.
Band, D., *et al.*: 1993, *Astrophys. J.* **413**, 281.
Baring, M. G.: 1989, *Astron. Astrophys.* **225**, 260.
Baring, M. G.: 1990, *Monthly Not. Roy. Astr. Soc.* **244**, 49.
Baring, M. G.: 1993, *Astrophys. J.* **418**, 391.

Baring, M. G.: 1995a, in Proc. of the NATO ASI *"Currents in High Energy Astrophysics"* eds. Shapiro, M. M., *et al.*, Kluwer, Dordrecht, p. 21.
Baring, M. G.: 1995b, *Adv. Space Res.* **15(5)**, 85.
Brainerd, J. J.: 1992, *Astrophys. J.* **384**, 545.
Brainerd, J. J.: 1994, *Astrophys. J.* **428**, 21.
Brainerd, J. J. and Lamb, D. Q.: 1987, *Astrophys. J.* **313**, 231.
Bussard, R. W. and Lamb, D. Q.: 1982, in *"Gamma-Ray Transients and Related Astrophysical Phenomena"* eds. Lingenfelter, R. E., *et al.*, AIP 77, New York, p. 189.
Carrigan, B. J. and Katz, J. I.: 1992, *Astrophys. J.* **399**, 100.
Daugherty, J. K. and Bussard, R. W.: 1980, *Astrophys. J.* **238**, 296.
Deguchi, S. and Watson, W. D.: 1986, *Astrophys. J.* **307**, 30.
Dermer, C. D.: 1989, *Astrophys. J.* **347**, L13.
Dermer, C. D.: 1990, *Astrophys. J.* **360**, 197.
Eichler, D. *et al.*: 1989, *Nature* **340**, 126.
Fenimore, E. E., *et al.*: 1988, *Astrophys. J. (Lett.)* **335**, L71.
Goodman, J.: 1986, *Astrophys. J. (Lett.)* **308**, L47.
Gould, A.: 1992, *Astrophys. J. (Lett.)* **386**, L5.
Graziani, C. *et al.*: 1992, in *"Gamma-Ray Bursts: Observations, Analyses and Theories"* eds. Ho, C., *et al.*, CUP, Cambridge, p. 407.
Harding, A. K.: 1991, *Phys. Rep.* **206**, 327.
Harding, A. K.: 1994, in *"The Second Compton Symposium"* eds. Fichtel, C. E., *et al.*, AIP 304, New York, p. 30.
Harding, A. K. and Preece, R. D.: 1989, *Astrophys. J. (Lett.)* **338**, L21.
Ho, C., Epstein, R. I. and Fenimore, E. E.: 1992, in *"Gamma-Ray Bursts: Observations, Analyses and Theories"* eds. Ho, C., *et al.*, CUP, Cambridge, p. 297.
Hurley, K., *et al.*: 1994, *Nature* **372**, 652.
Kaminker, A. D., Pavlov, G. G. and Mamradze, P. G.: 1987, *Astr. Space Sci.* **138**, 1.
Krolik, J. H. and Pier, E. A.: 1991, *Astrophys. J.* **373**, 277.
Lamb, D. Q., Wang, J. C. L., and Wasserman, I. M.: 1990, *Astrophys. J.* **363**, 670.
Liang, E. P.: 1986, *Astrophys. J.* **304**, 682.
Liang, E. P.: 1994, in *"Gamma-Ray Bursts"* ed. Fishman, G., *et al.*, AIP 307, NY, p. 351.
Liang, E. P., Jernigan, T. E. and Rodrigues, R.: 1983, *Astrophys. J.* **271**, 766.
Mazets, E. P., *et al.*: 1981, *Nature* **290**, 378.
Mazets, E. P.: 1995, to appear in *"High Velocity Neutron Stars and Gamma-Ray Bursts"* eds. Rothschild, R., *et al.*, AIP, New York.
Mészáros, P. and Rees, M. J.: 1993a, *Astrophys. J.* **405**, 278.
Mészáros, P. and Rees, M. J.: 1993b, *Astrophys. J. (Lett.)* **418**, L59.
Mészáros, P., Rees, M. J. and Papathanassiou, H.: 1994, *Astrophys. J.* **432**, 181.
Miller, G. S., *et al.*: 1991, *Phys. Rev. Lett.* **66**, 1395.
Mitrofanov, I. G.: 1984, *Astrophys. Space Sci.* **104**, 245.
Murakami, T., *et al.*: 1988, *Nature* **335**, 234.
Nemiroff, R. J., *et al.*: 1993, *Astrophys. J.* **414**, 36.
Paczyński, B.: 1986, *Astrophys. J. (Lett.)* **308**, L43.
Palmer, D. M., *et al.*: 1994, *Astrophys. J. (Lett.)* **433**, L77.
Preece, R. D. and Harding, A. K.: 1989, *Astrophys. J.* **347**, 1128.
Schneid, E. J., *et al.*: 1992, *Astron. Astrophys.* **255**, L13.
Shemi, A. and Piran, T.: 1990, *Astrophys. J. (Lett.)* **365**, L55.
Sommer, M., *et al.*: 1994, *Astrophys. J. (Lett.)* **422**, L63.
Stanek, K. Z., Paczyński, B. and Goodman, J.: 1993, *Astrophys. J. (Lett.)* **413**, L7.
Thompson, C.: 1994, *Monthly Not. Roy. Astr. Soc.* **270**, 480.
Ventura, J., Nagel, W. and Mészáros, P.: 1979, *Astrophys. J. (Lett.)* **233**, L125.
Wang, J. C. L., *et al.*: 1989, *Phys. Rev. Lett.* **63**, 1550.
Woosley, S. E.: 1993, *Astrophys. J.* **405**, 273.
Yoshida, A., *et al.*: 1991, *Publ. Astron. Soc. Japan* **43**, L69.
Zdziarski, A. A., Coppi, P. S. and Lamb, D. Q.: 1990, *Astrophys. J.* **357**, 149.